\documentclass[sn-mathphys]{sn-jnl}

\usepackage{caption}
\usepackage[table]{xcolor}

\jyear{2021}%

\DeclareMathOperator{\corr}{corr}
\DeclareMathOperator*{\argmax}{argmax}

\newcommand{\dataset}[2]{%
  \noalign{\vskip 3pt}%
  \rowcolor{gray!15}%
  \multicolumn{#1}{l}{\textit{#2}}\\
  \noalign{\vskip 3pt}%
}

\begin{document}

\title[ ]{Machine Intelligence on the Edge: Interpretable Cardiac Pattern Localisation Using Reinforcement Learning}

\author[1]{\fnm{Haozhe} \sur{Tian}}
\equalcont{These authors contributed equally to this work.}

\author[2]{\fnm{Qiyu} \sur{Rao}}
\equalcont{These authors contributed equally to this work.}

\author[3]{\fnm{Nina} \sur{Moutonnet}}

\author[1]{\fnm{Pietro} \sur{Ferraro}}

\author[2]{\fnm{Danilo} \sur{Mandic}}

\affil[1]{\orgdiv{Dyson School of Design Engineering}, \orgname{Imperial College London}, \orgaddress{\city{London} \postcode{SW7 2DB},  \country{UK}}}

\affil[2]{\orgdiv{Department of Electrical and Electronic Engineering}, \orgname{Imperial College London}, \orgaddress{\city{London} \postcode{SW7 2AZ},  \country{UK}}}

\affil[3]{\orgdiv{Department of Computing}, \orgname{Imperial College London}, \orgaddress{\city{London} \postcode{SW7 2RH},  \country{UK}}}

\abstract{
Matched filters are widely used to localise signal patterns due to their high efficiency and interpretability. However, their effectiveness deteriorates for low signal-to-noise ratio (SNR) signals, such as those recorded on edge devices, where prominent noise patterns can closely resemble the target within the limited length of the filter. One example is the ear-electrocardiogram (ear-ECG), where the cardiac signal is attenuated and heavily corrupted by artefacts. To address this, we propose the Sequential Matched Filter (SMF), a paradigm that replaces the conventional single matched filter with a sequence of filters designed by a Reinforcement Learning agent. By formulating filter design as a sequential decision-making process, SMF adaptively designs signal-specific filter sequences that remain fully interpretable by revealing key patterns driving the decision-making. The proposed SMF framework has strong potential for reliable and interpretable clinical decision support, as demonstrated by its state-of-the-art R-peak detection and physiological state classification performance on three challenging real-world ECG datasets. The proposed formulation can also be extended to a broad range of applications that require accurate pattern localisation from noise-corrupted signals.
}


\maketitle

\section{Introduction}
\label{sec:introduction}
\begin{figure}[h]
\centerline{\includegraphics[width=84mm]{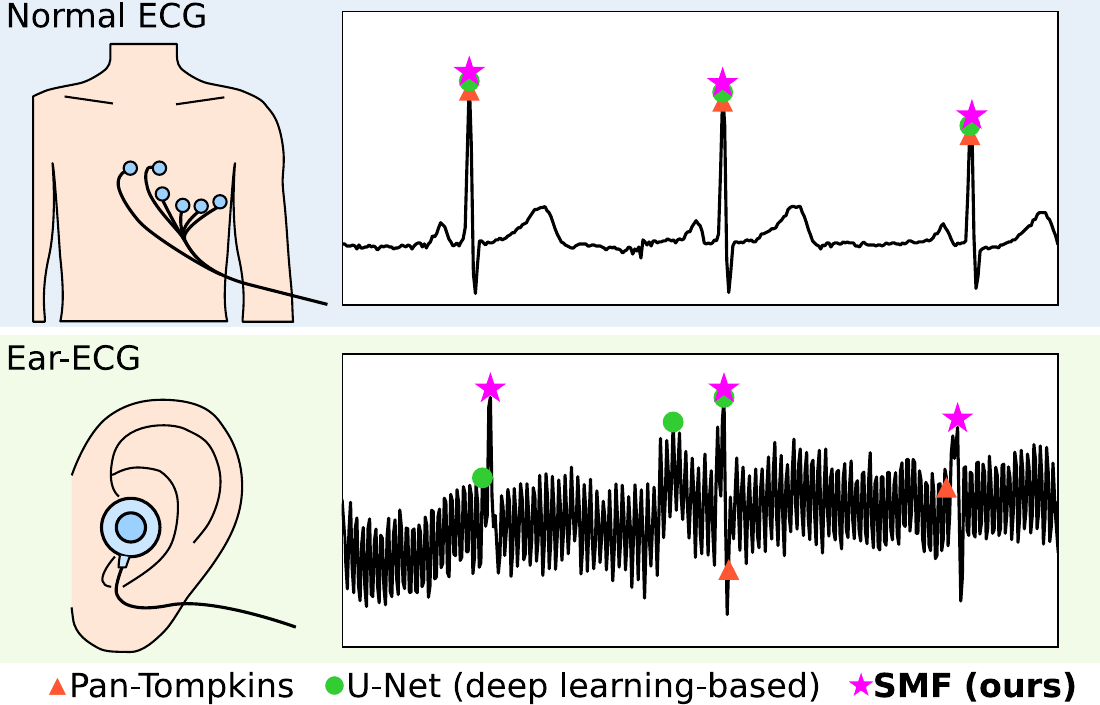}}
\caption{Despite the clear advantages in convenience, the critically low SNR of ear-ECG leads to poor R-peak detection performance by both the widely used Pan–Tompkins algorithm~\cite{pan1985real} and the state-of-the-art method using the U-Net~\cite{zahid2021robust}. In contrast, our proposed method, SMF, exhibits robust localisation performance on both normal ECG and the challenging ear-ECG.}
\label{fignormalear}
\end{figure}

Pattern localisation is a fundamental problem in signal processing, with applications spanning biomedicine, radar, and finance~\cite{fu2011review, jensen2012mining, li2018identification, chen2022joint}. The recent development of edge signal acquisition devices has created a strong demand for more robust pattern localisation methods capable of tackling low signal-to-noise ratio (SNR) conditions. One prominent example of pattern localisation on edge is detecting (localising) R-peaks—the patterns corresponding to the electrical activity of ventricular depolarisation—in ear-electrocardiograms (ECGs), which are recorded via electrodes placed in the ear canal instead of on the chest~\cite{xu2024earable}. Whilst being more convenient to set up and record over a prolonged period of time, the ear-ECG signal suffers from an extremely low SNR in comparison to the chest ECG. This is due to the attenuation of the ECG signal during propagation to the ear canal, plus the ear’s proximity to non-cardiac sources of noise, including major arteries, muscles, and brain activity. Consequently, despite its central role in advanced physiological monitoring~\cite{thayer2012meta, kleiger2005heart, cole1999heart, wang2023ecg, gendler2025heart}, R-peak localisation in ear-ECG remains challenging, with existing methods failing to deliver reliable performance (see Fig.~\ref{fignormalear}).

Outside of ear-ECG, advanced pattern localisation methods have been applied to detect R-peaks in low-SNR ECG signals. These methods often rely on deep learning~(DL), where neural networks are trained to minimise a proxy loss function, such as the Binary Cross-Entropy (BCE) loss, so that they predict whether each sample in an ECG segment corresponds to an R-peak~\cite{laitala2020robust, zhou2020deep, peng2023ecg}. The recent advances in neural network architectures, such as the U-Net~\cite{ronneberger2015u}, have further improved the performance of DL-based R-peak detection methods~\cite{zahid2021robust}. However, these approaches suffer from poor interpretability, as it is unclear which signal patterns drive the localisation decisions. Moreover, optimising the proxy loss function does not fully align with clinically relevant performance metrics, such as the true positive, false positive, and false negative rates of R-peak detection.

An alternative, more interpretable approach is the matched filter (MF)~\cite{hamilton1988adaptive, xue1992neural}, which exploits the characteristic QRS pattern around R-peaks. After correlating a QRS-like template with a noisy ECG signal, high correlation values appear at the target QRS pattern locations (i.e., the R-peaks), whereas low values occur elsewhere due to morphological differences between the template and noise~\cite{chanwimalueang2015enabling}. However, MF is inherently limited in differentiating true R-peaks from artefacts with high prominence and similar morphology. This challenge is particularly acute in ear-ECG, where the amplitudes of artefact peaks are comparable to or even exceed the target R-peaks. Moreover, existing MF templates are typically manually defined or derived from historical QRS patterns. Although recent work has explored learning the MF template~\cite{davies2024deep}, the resulting template remains static at deployment, making it suboptimal for ECGs with non-stationary QRS patterns, such as those seen in arrhythmia patients~\cite{clifford2006advanced}.

In this work, we propose the Sequential Matched Filter (SMF), which replaces the conventional single-MF paradigm with a sequence of signal-specific filters to achieve robust performance while retaining full interpretability.
To automatically design the filter sequences, SMF performs sequence-level planning using a Reinforcement Learning (RL) agent, a data-driven approach that leverages recent advances in deep neural networks to achieve competitive performance in complex sequential decision-making tasks~\cite{silver2017mastering, ying2021optimal, degrave2022magnetic, aung2025real, chen2025target}. SMF offers several key advantages over existing pattern localisation methods:
\begin{itemize}
    \item SMF distinguishes the target pattern from prominent noise patterns with similar waveforms by iteratively exploiting subtle morphological differences.
    \item SMF adapts to non-stationary target patterns, such as those observed in arrhythmia patients, by tailoring MF templates for each signal segment.
    \item SMF offers full interpretability by revealing, at each step, the key signal patterns that inform the final localisation decision.
    \item SMF outperforms existing DL-based methods by directly optimising localisation performance metrics without a proxy loss function.
\end{itemize}

To the best of our knowledge, SMF is the first method to automate sequentially applied MFs for pattern localisation. Despite using an edge-deployable neural network with only $\sim$150\,k parameters ($\sim$0.57\,MB), SMF significantly outperforms state-of-the-art DL baselines and non-sequential MFs optimized by the same RL algorithm, without sacrificing interpretability. The competitive performance of SMF is empirically verified on three challenging real-world datasets: (i) a noisy \emph{ear-ECG} dataset collected using our own custom-built in-ear sensors~\cite{goverdovsky2017hearables, tian2023hearables}, (ii) a pathological \emph{arrhythmia ECG} dataset recorded from subjects with atrial fibrillation using handheld edge devices~\cite{clifford2017af}, and (iii) the large-scale \emph{Icentia11k} dataset consisting of continuous daily ECG recordings acquired via wearable chest patches~\cite{PhysioNet-icentia11k-continuous-ecg-1.0}.
Additionally, from the detected R-peaks, we derive Heart Rate Variability (HRV) features to classify physiological states, where SMF exhibits robust performance, highlighting its potential for real-world cardiac health monitoring. We believe that the RL agent-driven filter design paradigm presented in this paper extends beyond the biomedical signal processing domain to a wide range of signal processing tasks that demand lightweight yet robust pattern localisation.

\section{Preliminary}
\label{sectoyexp}

\begin{figure*}[t]
\vspace{1em}
\centerline{\includegraphics[width=\textwidth]{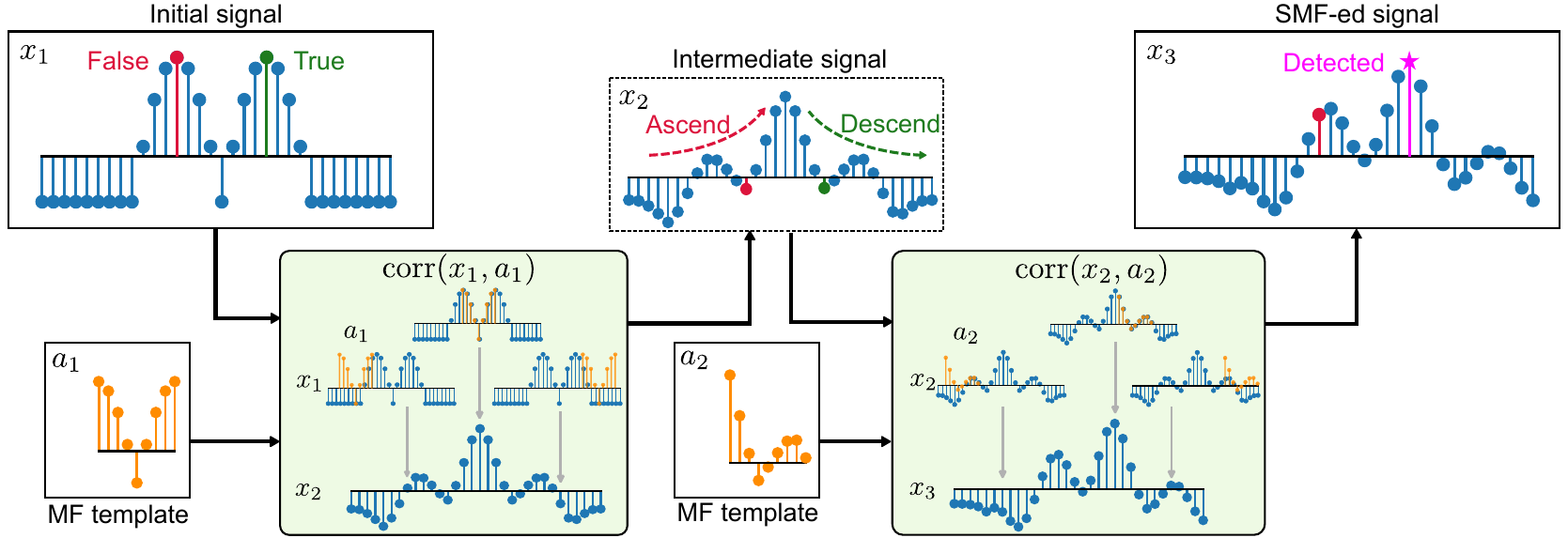}}
\vspace{20pt}
\caption{Overcoming the limitation of single-stage MF using a sequence of MFs. No single-stage MF can distinguish between the false and the true peaks in $x_1$, as the patterns surrounding them are identical within the span of the length-9 MF template. However, a sequence of only two MFs can accurately localise the true peak: the first MF, $a_1$, introduces pattern variation in $x_2$, 
and the second MF, $a_2$, correlates with the descending pattern that corresponds to the true peak.}
\label{figexp}
\end{figure*}

In this section, we describe the classic MF method and present an example illustrating its inherent limitation in distinguishing between similar patterns. For a signal $x_t\in\mathbb{R}^L$, applying a MF with template $a_t\in\mathbb{R}^H$ produces the filtered signal, $x_{t+1}\in\mathbb{R}^L$, whose $n$-th sample is calculated as
\begin{align}
    x_{t+1}(n) = \sum_{k=0}^{H-1} a_t(k) x_t(n+k-\left\lfloor\frac{H}{2}\right\rfloor),
\label{eqcorr}
\end{align}
where $a_t(i)=0$ when $i< 0$ or $i\geq H$, $x_t(i)=0$ when $i< 0$ or $i\geq L$, and the operator $\left\lfloor i\right\rfloor$ is the greatest integer less than or equal to $i$. Equation \eqref{eqcorr} can be interpreted as sliding the template $a_t$ over the signal $x_t$ and measuring their correlation at these positions (see illustrations in Fig.~\ref{figexp}, green blocks). Ideally, the correlation level, $x_{t+1}$, should peak at the index of the target pattern, allowing straightforward pattern localisation by locating the prominent maxima. However, MF performance degrades when noise contains patterns similar to the target, as illustrated in the example in Fig.~\ref{figexp}, where distinguishing the true peak (green, right) from the false peak (red, left) in $x_1$ is infeasible for any template $a_t$ with limited length ($H=9$ in this case), since the patterns surrounding both peaks are identical within the span of the template. Although a template can be designed to match the transition from the true peak to the negative baseline, the resulting correlation maxima would be delayed, introducing a temporal shift in the identified peak location.

On the other hand, the true peak can be localised with a sequence of merely two strategically designed MFs. As noted previously, a single-stage MF cannot distinguish between the false and true peaks in Fig.~\ref{figexp} because both appear identical within the span of the filter. However, by designing a template $a_1$ that matches the valley shape between the two peaks, the resulting output $x_2$ ascends around the false peak and descends around the true peak, thereby producing distinctive pattern variations between the noise and the target. A second template, $a_2$, is then shaped to match the descending pattern of $x_2$ near the true peak, enabling accurate extraction of the true peak. Consequently, $x_3$ exhibits its largest amplitude at the time index of the true peak. As illustrated in Fig.~\ref{figexp}, the overall output $x_3$ indeed peaks at the true peak location, confirming the successful localisation of the true peak.

\section{Methodology}
\label{secmeth}

\subsection{SMF as a Sequential Decision-Making Process}
\begin{figure*}[t]
\vspace{1em}
\centerline{\includegraphics[width=\textwidth]{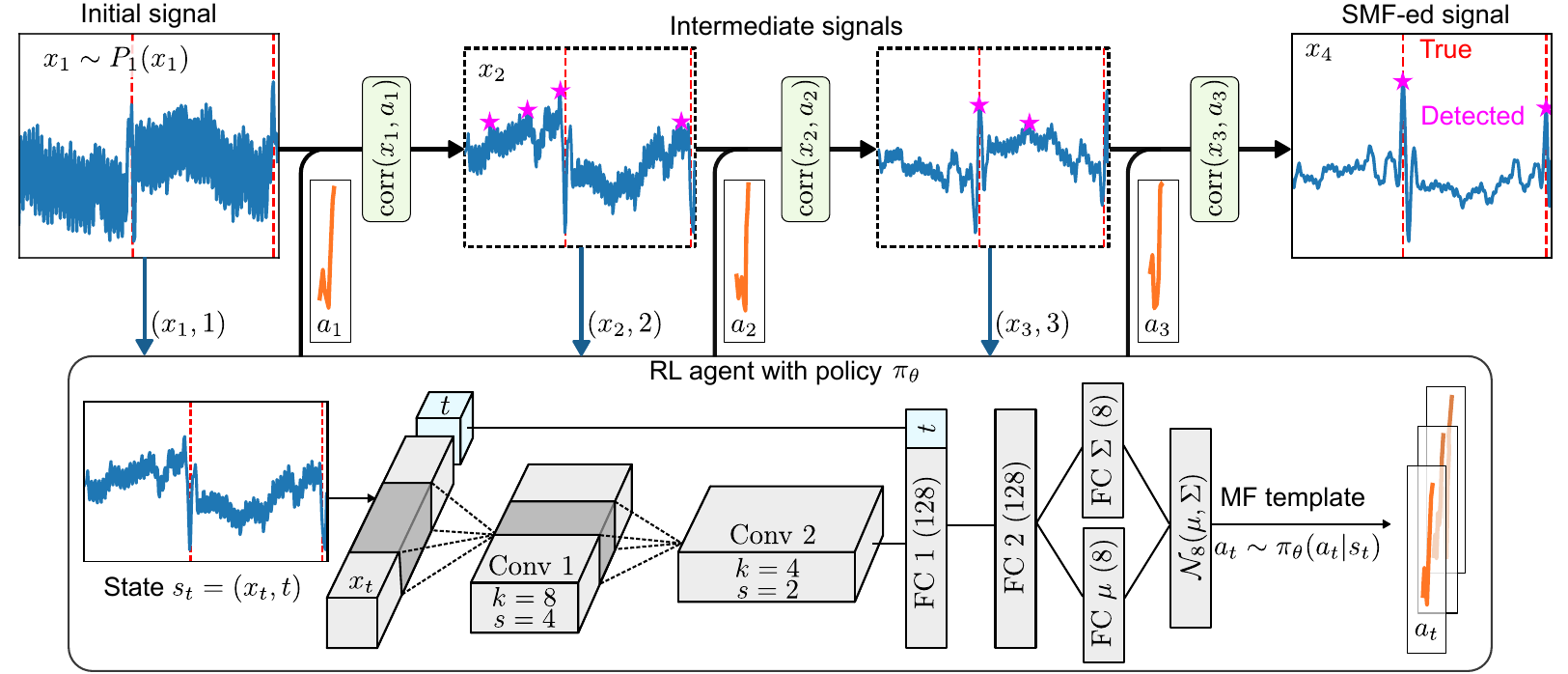}}
\caption{The workflow of the SMF for R-peak detection. The initial signal $x_1$ is sampled from the ECG dataset with distribution $P_1(x)$. At time step $t$, the state $s_t = (x_t, t)$ contains the signal $x_t$ and the time step $t$, based on which the RL agent generates the MF template, $a_t$, for calculating $x_{t+1}$, which is iteratively used as the state for step $t+1$. The RL agent has policy $\pi_\theta$ in the form of a neural network that generates stochastic $a_t\sim\pi_\theta(a_t\mid s_t)$. After training, SMF templates $a_t$ are interpretable, as they reveal key signal patterns at each step.}
\label{figsche}
\end{figure*}

The example in Section~\ref{sectoyexp} illustrates the effectiveness of strategically designing and iteratively applying sequence-level optimised MFs. To this end, we model SMF as a sequential decision-making process involving interactions between two key components: an \textit{environment} and an \textit{agent}. An \textit{episode} consists of $N$ cascaded MF steps, where the output of each step becomes the input to the subsequent step, as illustrated in Fig.~\ref{figsche} for $N=4$. In the first step, the environment is initialised with an ECG segment that is randomly sampled from the training set during training, or taken directly from real-time ECG data during deployment. At each step, the RL agent takes in the environment signal (state) and generates an MF template (action), which is correlated with the environment signal. The correlation output replaces the previous environment signal. In the next step, the same procedures are repeated. After the last ($N$-th) step, R-peaks are localised from the environment's stored signal by identifying local maxima that exceed a threshold of 0.5 and are separated by at least 30 samples. During training, these localised R-peaks are compared with the ground-truth R-peaks to calculate a performance metric, which is used as a reward signal that informs the learning of the RL agent. During testing or deployment, where ground-truth R-peaks are unavailable, SMF automatically detects the R-peaks without external guidance.

Formally, the SMF framework is a Markov Decision Process (MDP), because the next state, i.e., the MF correlation output, is determined only by the current signal and the MF template applied. We denote the MDP as $\mathcal{M}=(\mathcal{X}, \mathcal{S}, \mathcal{A}, r, \corr)$. The signal space $\mathcal{X}\subseteq\mathbb{R}^L$ contains both the original and SMF-transformed ear-ECG signals. In this work, we set the length of collected ECG segments to $L=250$. The state space is a set $\mathcal{S} = \{s_t=(x_t, t) \mid  x_t\in\mathcal{X}, t \in \{1, \dots, N\}\}$ that contains states $s_t$. Each $s_t\in\mathcal{S}$ is a tuple of the normalized signal $x_t$ and its corresponding time step $t$ within an episode. The action space is a set $\mathcal{A}\subseteq\mathbb{R}^H$ that contains all possible MF templates. We set $H=8$, which is a short enough length for most edge applications. We define the reward function $r:\mathcal{S}\rightarrow\mathbb{R}$ as
\begin{align}
\begin{split}
    r(s_t) &= \delta_{tN} f(\mathrm{TP, FP, FN}) \\
    &= \delta_{tN} (10\mathrm{TP} - 5\mathrm{FP} - 5\mathrm{FN}),
\end{split}
\label{eqreward}
\end{align}
where the $\delta_{ab}=1$ if $a=b$ and $\delta_{ab}=0$ if $a\neq b$. 
The final peak locations were obtained by applying the SciPy function find\_peaks()~\cite{2020SciPy-NMeth} to the last state $s_N$ of the episode. Specifically, we set the height threshold to 0.5 to suppress persistent low-amplitude fluctuations in $s_N$, and enforced a minimum peak distance of 30 samples (approximately 120~ms at 250~Hz), which covers the typical duration of a QRS complex, to avoid multiple detections within a single heartbeat. The detected peaks were matched to ground-truth R-peaks using a tolerance of 5 time steps (20~ms), which is within the commonly used range in standard ECG benchmarks~\cite{liu2018performance}.
The true positive (TP) is the number of correctly identified R-peaks, the false positive (FP) is the number of falsely identified peaks, and the false negative (FN) is the number of missed peaks.
The $10\mathrm{TP}$ term is used in \eqref{eqreward} to prioritize correct heartbeat detections. The penalty terms $-5\mathrm{FP}$ and $-5\mathrm{FN}$ in \eqref{eqreward} are introduced for two reasons: (i) they symmetrically encourage both high precision and high recall, since falsely identified peaks and missed peaks degrade practical downstream metrics (e.g, heart rate and HRV) at roughly comparable scales; (ii) they provide a meaningful learning signal during early training. If only TPs are rewarded, a weak initial policy, which rarely detects correct R-peaks, would almost always receive zero reward and thus learn very slowly. Empirically, we observed consistent performance when the FP and FN penalty weights were $-3$, $-5$, and $-7$, indicating that the learned policy is not sensitive to the exact coefficient values.
The correlation function $\corr: \mathcal{X}\times\mathcal{A}\rightarrow\mathcal{X}$ is defined as $x_{t+1} = \corr(x_t, a_t)$, where the $n$-th sample of $x_{t+1}$ is calculated using \eqref{eqcorr}.
To prevent uncontrolled scaling effects, the template, $a_t$, is constrained to the same range as $x_t$, and the correlation output, $x_{t+1}$, is renormalized at each SMF step. Although we deviate from the standard MDP definition by introducing an additional internal signal space $\mathcal{X}$, the expression in \eqref{eqcorr} still ensures the core Markov property, i.e., the upcoming state of SMF depends only on the current state and the current action.

We define the stochastic SMF policy as $\pi:\mathcal{S}\times\mathcal{A}\rightarrow\mathbb{R}$, which takes an state $s_t$ as input and outputs a distribution over MF templates, $\pi(a_t\mid s_t)$. 
The stochastic $\pi$ explores various MF template sequences during training, thus avoiding getting stuck with suboptimal solutions.
The optimisation objective is to maximise the expected cumulative reward in an episode, that is
\begin{align}
    \pi^\star = \argmax_\pi \mathbb{E}_{s_1, a_1, \cdots}\left[\sum_{t=1}^N r(s_t)\right],
\label{eqrlobj}
\end{align}
where $s_t = (x_t, t)$, $x_1$ is sampled from the ECG dataset with $x_1\sim P_1(x_1)$, $x_{t+1}=\corr(x_t, a_t)$, and $a_t\sim\pi(a_t\mid s_t)$. For the value function, $V^\pi$, and the state-action value function, $Q^\pi$, which will be used later for training the policy $\pi$, we follow the standard definitions from~\cite{pmlr-v37-schulman15}, given by:
\begingroup
\allowdisplaybreaks
\begin{align}
    &V^\pi(s_t) = \mathbb{E}_{a_t, s_{t+1}, a_{t+1}, \dots}\left[\sum_{i=t}^N r(s_i)\right], \notag \\
    &Q^\pi(s_t, a_t) = \mathbb{E}_{s_{t+1}, a_{t+1}, \dots}\left[\sum_{i=t}^N r(s_i)\right]. \label{eqqv}
\end{align}
\endgroup

\subsection{Optimising SMF with RL}
\label{secRLmethodology}

\begin{algorithm}[t]
\vspace{1em}
   \caption{Training SMF}
   \label{alg:SMF}
\begin{algorithmic}[1]
    \State {\bfseries Initialization:} Train set $\mathcal{X}_\text{train}$ containing ECG segments and ground-truth R-peak positions, episode length $N$
    \State {\bfseries Output:} Trained SMF policy $\pi_\theta$
    \Repeat
    \State Randomly sample $(x_1, \{\text{peaks}\}) \in \mathcal{X}_\text{train}$
    \State $t \gets 1$
    \While{$t \leq N$}
    \State Get MF template: $a_t \sim \pi_\theta(a_t \mid  s_t = (x_t, t))$
    \State $x_{t+1}(n) = \sum_{k=0}^{H-1} a_t(k) x_t(n+k-\left\lfloor\frac{H}{2}\right\rfloor)$
    \If{$t == N$}
    \State Find local maximums $\{\text{preds}\}$.
    \State Compute TP, FP, and FN by comparing $\{\text{preds}\}$ and $\{\text{peaks}\}$
    \State $r(x_t) = 10\mathrm{TP} - 5\mathrm{FP} - 5\mathrm{FN}$
    \Else
    \State $r(x_t) = 0$
    \EndIf
    \State $s_{t+1}=(x_{t+1}, t+1)$
    \State Use $(s_t, a_t, s_{t+1}, r(s_t))$ to update $\pi_\theta$ (e.g., using PPO or SAC).
    \State $t = t+1$
    \EndWhile
    \Until{$convergence$ is $true$}
\end{algorithmic}
\end{algorithm}

To address the continuous state space $\mathcal{S}$ and action space $\mathcal{A}$ in SMF, this study employs deep RL, where the policy $\pi$ is approximated using a neural network $\pi_\theta$ parametrized by $\theta$ to allow generalization to unobserved states and actions in $\mathcal{S}$ and $\mathcal{A}$. The policy $\pi_\theta$ takes in a state tuple $s_t = (x_t, t)$ of size $250+1$, and outputs an MF template of length $8$, using the neural network illustrated in the lower panel of Fig.~\ref{figsche}. The input state $s_t = (x_t, t)$ comprises a signal $x_t$ of length $L=250$ and a scalar time index $t$. 
To facilitate neural network training, both $x_t$ and $t$ were normalized to the range $[-1, 1]$.
The signal $x_t$ is processed by two 1-dimensional Convolutional Neural Network (CNN) layers (Conv): the first with a kernel size of $k=8$ and stride $s=4$, and the second with a kernel size of $k=4$ and stride $s=2$. The convolution output is then transformed into a feature vector of size 128 using a fully connected layer (FC). This feature vector is concatenated with the scalar, $t$, resulting in a combined representation of size 129. This concatenated representation is further transformed by an FC layer into a vector of size 128, which is subsequently mapped through two separate FC layers to generate the mean vector $\mu$ (size 8) and the diagonal covariance matrix $\mathbf{\Sigma} = \mathrm{diag}(\sigma_1, \dots, \sigma_8)$. Together, $\mu$ and $\mathbf{\Sigma}$ define a multivariate Gaussian distribution $\mathcal{N}_8(\mu, \mathbf{\Sigma})$, from which the MF template, $a_t$, is sampled using the reparametrization trick~\cite{haarnoja2018soft}.

The general procedure for training $\pi_\theta$ is summarised in Algorithm \ref{alg:SMF}. Most existing deep RL algorithms can be used to solve the optimisation problem in \eqref{eqrlobj}, among which the two most prominent categories are: i) policy gradient methods, which directly optimise $\pi_\theta$ using the objective in \eqref{eqrlobj}; and ii) actor–critic methods, which optimise $\pi_\theta$ through estimated state–action values, enabling learning from past experience and thereby improving sample efficiency. This work proposes two SMF implementations based on two state-of-the-art RL algorithms, each representing one of the two abovementioned categories.

\subsubsection{SMF-PPO}
\label{secppometh}

Proximal Policy Optimisation (PPO) is a prominent policy gradient method that clips the objective to prevent excessive policy updates, thereby stabilising training~\cite{schulman2017proximal}. The SMF-PPO first collects a batch of SMF episodes using $\pi_{\theta_\text{old}}$, then performs update by maximising the following objective, as
\begin{align}
    &\max_\theta\hat{\mathbb{E}}_{s_t, a_t}\left[ L(s_t, a_t, \theta_{\text{old}}, \theta)\right]\text{, where} \notag \\
    &L(s_t, a_t, \theta_{\text{old}}, \theta) = \left\{\begin{matrix}
    \min \left(\frac{\pi_\theta(a_t\mid s_t)}{\pi_{\theta_{\text{old}}}(a_t\mid s_t)}, 1+\epsilon\right)\hat{A}_t, &  \hat{A}_t > 0\\
    \max \left(\frac{\pi_\theta(a_t\mid s_t)}{\pi_{\theta_{\text{old}}}(a_t\mid s_t)}, 1-\epsilon\right)\hat{A}_t, &  \hat{A}_t \leq 0\\
    \end{matrix}\right., \notag \\
    &\hat{A}_t = r(s_N) - V_\psi(s_t).
\label{eqPPOloss}
\end{align}
At state $s_t$, $\hat{A}_t$ captures how much better (or worse) each chosen MF template, $a_t$, performed compared to the expected average. The performance of $a_t$ is evaluated as the final peak extraction performance $r(s_N)$ in the episode that includes $s_t$. The expected average is estimated using the value function, $V^\pi$, defined in \eqref{eqqv}. A positive $\hat{A}_t$ indicates that MF template $a_t$ outperforms the average of $\pi_{\theta_\text{old}}$. Therefore, to maximize $L(s_t, a_t, \theta_{\text{old}}, \theta)$, it is favourable to increase the probability of using $a_t$, i.e., increase $\pi_\theta(a_t\mid s_t)$. However, the $\min(\cdot)$ ensures that the updated $\pi_\theta$ is not too far from the old $\pi_{\theta_{\text{old}}}$, i.e., $\pi_\theta(a_t\mid s_t) \leq (1+\epsilon)\pi_{\theta_{\text{old}}}(a_t\mid s_t)$. The same logic applies to the case where $\hat{A}_t$ is negative. The clipping in SMF-PPO avoids drastic updates on MF templates, which could result in catastrophic performance degradations.

In practice, $V^\pi$ is approximated by a neural network $V_\psi$, which shares a similar architecture with the policy network $\pi_\theta$ but outputs a scalar value. The parameters $\psi$ are updated by minimising the mean squared error between $V_\psi$ and its Bellman estimation
\begin{align*}
    \min_\psi \hat{\mathbb{E}}_{s_t, s_{t+1}}\left[(V_\psi(s_t) - (r(s_t)+V_\psi(s_{t+1}))^2\right].
\end{align*}
In SMF-PPO, the estimation of the advantage $\hat{A}_t$ in \eqref{eqPPOloss} is further stabilised using the Generalised Advantage Estimator (GAE)~\cite{schulman2015high}.

\subsubsection{SMF-SAC}
\label{secsacmeth}

Soft Actor-Critic (SAC) is a prominent actor-critic method that uses an entropy regularisation term in its objective to encourage diverse and robust policies~\cite{haarnoja2018soft}. The SMF-SAC updates its policy at every SMF step using a sampled batch of historical MF steps, each denoted by $(s_t, a_t, s_{t+1}, r(s_t))$. SMF-SAC parametrises the state-action value function $Q^\pi$ with a neural network, $Q_\phi$, with parameter $\phi$, to allow generalisation to unobserved state-action pairs. The network structure of $Q_\phi$ is similar to the policy network $\pi_\theta$ but outputs a scalar value. The $\phi$ is updated by minimizing the mean square error between $Q_\phi(s_t, a_t)$ and its Bellman estimator $r(s_t) + Q_\phi(s_{t+1}, a_{t+1})$
\begin{align*}
    &\min_\phi \hat{\mathbb{E}}_{s_t, a_t, s_{t+1}}\left[(Q_\phi(s_t, a_t) - y)^2\right]\text{, where} \notag\\
    &y=r(s_t) + Q_\phi(s_{t+1}, a_{t+1}) - \alpha \log \pi(a_{t+1}\mid s_{t+1}), \notag\\
    &a_{t+1}\sim\pi(a_{t+1}\mid s_{t+1}),
\end{align*}
where the entropy regularisation term $\log \pi(a_{t+1}\mid s_{t+1})$ introduced by~\cite{haarnoja2018soft} encourages the exploration of diverse MF templates and avoids sticking to sub-optimal templates. The hyperparameter $\alpha$ adjusts the intensity of regularisation.

To update the actor $\pi_\theta$, observe that the estimated state-action value $Q_\phi$ indicates the expected future cumulative rewards of taking $a_t\sim\pi_\theta(a_t\mid s_t)$ at $s_t$, which the updated policy aims to maximise. Therefore, the $\pi_\theta$ is directly updated by maximizing the expected value of $Q_\phi$, as 
\begin{align*}
    \max_\theta \hat{\mathbb{E}}_{s_t}\left[ Q_\phi(s_t, a_t) - \alpha \log \pi(a_{t+1}\mid s_{t+1})\right], a_t \sim \pi_\theta(a_t\mid s_t).
\end{align*}

\section{Hardware and Data Acquisition}
\label{secSetup}

\begin{figure}[t]
\vspace{1em}
\centerline{\includegraphics[width=84mm]{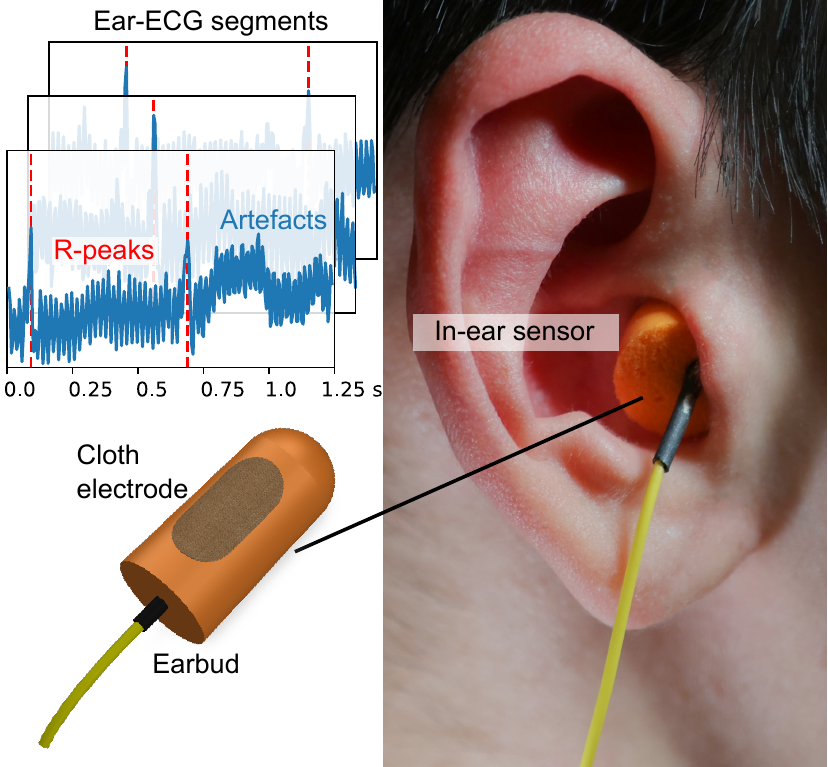}}
\caption{The ear-ECG acquisition setup. The ear-ECG signal was recorded using our custom-built in-ear sensors, with the right ear used for recording and the left ear serving as the reference. The in-ear sensors consist of an earbud and a soft cloth electrode. The ear-ECG signals were recorded at a sampling rate of 200~Hz and were split into segments with 250 samples (1.25s).}
\label{figsetup}
\end{figure}

The ear-ECG dataset in this study was recorded from 7 healthy subjects (5 males and 2 females, aged 20-30) under the ethics protocol JRCO 20IC6414. Our custom-built in-ear sensors were placed in both ear canals, recording signals from the right ear while using the left ear as the reference. Fig.~\ref{figsetup} depicts the in-ear sensor based on the design described in~\cite{goverdovsky2017hearables}, which consists of an earbud and a soft cloth electrode. The earbud was made of viscoelastic foam to alleviate artefacts arising from mechanical deformations of the ear canal. The cloth electrode was made of stretchable, low-impedance fabric, ensuring reliable skin contact and enhanced comfort during prolonged wear. To further reduce impedance caused by poor skin contact, conductive gel was applied before placing the sensor in the ear canal. The signal was recorded at a sampling rate of 200~Hz and was subsequently divided into non-overlapping 250-sample segments (1.25s). The left panel of Fig.~\ref{figsetup} presents some sample ear-ECG segments recorded using our setup. The identification of true R-peaks is extremely challenging in these ear-ECG segments due to the numerous false artefact peaks around R-peaks. The prominent false peaks with similar or even greater amplitude compared to true R-peaks pose significant challenges for accurate R-peak detection.

\section{Experiment Setup}
\label{secexpset}

\subsection{Datasets}

We validated the SMF\footnote{Code available at \url{https://github.com/HaozheTian/Sequential-Matched-Filter}} on three real-world ECG datasets:
\begin{enumerate}
  \item The \textit{ear-ECG} dataset contains 720 ear-ECG segments collected using the setup described in Section~\ref{secSetup}. The Hearable setup led to low signal amplitude and prominent artefact peaks, making accurate R-peak detection particularly challenging (see example in Fig.~\ref{figvis}a).
  \item The \textit{arrhythmia ECG} dataset contains 771 single-lead ECG recordings derived from subjects with atrial fibrillation, the most common cardiac arrhythmia, in the 2017 Computing in Cardiology Challenge~\cite{clifford2017af}. This dataset is challenging because arrhythmia causes non-stationary QRS patterns and irregular R-R intervals (the distance between successive R-peaks), as illustrated in Fig.~\ref{figvis}b. Additionally, the arrhythmia ECG dataset, recorded using the handheld AliveCor device, is relatively noisy and susceptible to electrode misplacement, which can invert the signal in some segments (see the left panel of Fig.~\ref{figvis}c).
  \item The \textit{Icentia11k} dataset is a large-scale wearable ECG database collected using the CardioSTAT device, a single-lead chest-mounted patch~\cite{PhysioNet-icentia11k-continuous-ecg-1.0}. The recordings contain substantial motion artefacts as they were continuously acquired during subjects' daily activities. From this dataset, we extracted 20,000 ECG segments for our experiments.
\end{enumerate}
Since the arrhythmia ECG and Icentia11k datasets were originally sampled at different sampling rates, we resampled them to 200~Hz so that each 250-sample segment corresponds to 1.25~s of recording, consistent with the ear-ECG dataset.
We used 70\% of the ECG segments for training and the remaining 30\% for testing. The train-test split remained consistent across all experiments. The test set was strictly reserved for evaluation, ensuring that neither the SMF nor the baseline methods could access it during training. To train the SMF, we designed three RL environments in the widely used OpenAI Gym style~\cite{brockman2016openaigym}, corresponding to the ear-ECG, arrhythmia ECG, and Icentia11k datasets. During training, the RL environments were reset every $N$ steps with a randomly selected ECG segment from the training set. During testing, the $N$-step SMF was applied to all ECG segments in the test set for calculating the average performance metrics.

\subsection{Implementations of SMF and Baselines}

To train the proposed SMF method, $10^{5}$ SMF steps were observed. For SMF-PPO, the policy $\pi_\theta$ was updated after every $500$ consecutive transitions. For each transition, the cumulative reward used for advantage estimation was the final R-peak detection performance of the same episode. The 500 collected single-step transitions were then randomly split into four mini-batches of 125 transitions. The policy update was performed over four epochs, with each epoch iterating through all mini-batches. The learning rate was $10^{-4}$ and the clipping ratio $\epsilon$ was set to $0.2$. Gradient clipping was applied to avoid gradients larger than $0.5$. For SMF-SAC, the policy $\pi_\theta$ was updated every 2 steps using a batch of $512$ historical single-step transitions stored in a Replay Buffer~\cite{mnih2013playing}. A Polyak weight averaging with a smoothing factor $0.005$ was used to stabilise the update of $Q$ networks~\cite{mnih2015human}. The learning rate was set to $10^{-4}$ for both policy updates and $Q$ network updates. The entropy regularisation term $\alpha$ was set to $0.2$. The training setup and hyperparameters were kept the same for all experiments in this section.

The neural network architectures used by SMF are lightweight and well-suited for edge deployment. For SMF-PPO, sharing parameters between the value function and the policy results in an RL agent with approximately 156\,k parameters ($\approx$0.60\,MB). For SMF-SAC, both the $Q$-network and the policy network contain around 140\,k parameters ($\approx$0.54\,MB). As demonstrated in our previous work~\cite{zylinski2023hearables}, these RL agents can be readily deployed on edge devices to achieve real-time pattern localisation, requiring only milliseconds to process 60-second ECG recordings on an Android smartphone.

To evaluate the performance and robustness of SMF, we empirically compared it against the following baselines.
\begin{enumerate}
    \item The \textit{Pan-Tompkins} algorithm~\cite{pan1985real} is arguably the most widely used R-peak detection method that combines a series of complex signal processing operations and a decision rule that decides the validity of each potential R-peak by comparing the current R-R interval to the average of historical R-R intervals.
    \item The \textit{Bidirectional RNN (Bi-RNN)}~\cite{schuster1997bidirectional} is a popular sequential neural network architecture that uses both forward and backward recurrent layers to capture context from past and future signals.
    \item The \textit{U-Net}~\cite{ronneberger2015u, zahid2021robust} is a CNN-based neural network architecture that \cite{zahid2021robust} reports achieving state-of-the-art performance in ECG R-peak detection.
    \item The \textit{MF-PPO} and \textit{MF-SAC} are ablated versions of the proposed SMF algorithm, restricted to episode lengths of 1. We include them as baselines to represent the performance of single-stage MFs when optimised in a RL-driven manner. Although non-sequential, these baselines generate MF templates that directly optimise R-peak detection performance by using the RL reward function in \eqref{eqreward} as their objective.
\end{enumerate}
The Bi-RNN baseline consisted of two Bi-RNN layers with hidden sizes of 64, which transformed the signal into a feature vector of length 250 with 128 channels, followed by a linear layer that mapped this feature vector to a length 250 prediction vector.
The U-Net baseline followed~\cite{zahid2021robust}.
As discussed in Section~\ref{sec:introduction}, it is infeasible to directly optimise TP, FP, and FN with DL-based methods. Therefore, the objectives of the DL methods were to minimise the binary cross-entropy (BCE) loss between the network prediction vector and a binary vector of length 250, where ones correspond to R-peak positions and zeros to non-peak positions. To convert Bi-RNN and U-Net outputs into R-peak predictions, we followed the peak-picking procedures described in~\cite{zahid2021robust}. 
In addition to the height and distance thresholds (as used in SMF), this procedure applies additional verification steps that remove abnormal beat detections based on Pearson correlation. Both Bi-RNN and U-Net were trained over 1000 epochs with a batch size of 100 and a learning rate of $0.005$. Across all tests, the detected peaks were matched to ground-truth R-peaks using a tolerance of
5 time steps (20 ms).

\section{Results and Analysis}
\label{secres}

\subsection{Comparison of R-peak Detection Performance.}

\begin{table}[t]
\caption{R-peak detection performance.}
\setlength{\tabcolsep}{8pt}
\centering
\begin{tabular}{lccc}
\hline
\multicolumn{1}{c}{}       & Precision & Recall & F-1 score\\\hline
\dataset{4}{Ear-ECG}
\multicolumn{1}{p{2.5cm}}{Pan-Tompkins\hfill\cite{pan1985real}} &  0.5520   & 0.5243 & 0.5378    \\
\multicolumn{1}{p{2.5cm}}{Bi-RNN\hfill\cite{schuster1997bidirectional}}       &  0.7670   & 0.8778 & 0.8187    \\
\multicolumn{1}{p{2.5cm}}{U-Net\hfill\cite{zahid2021robust}}        &  0.9029   & 0.9490 &    0.9254\\
\multicolumn{1}{p{2.5cm}}{MF-PPO}    &  0.9900   & 0.9612 & 0.9754\\
\multicolumn{1}{p{2.5cm}}{MF-SAC}    &  0.9700   & 0.9417 & 0.9557\\
\multicolumn{1}{p{2.5cm}}{SMF-PPO}    &  0.9902   & 0.9806 & 0.9854\\
\multicolumn{1}{p{2.5cm}}{SMF-SAC}    &  \textbf{1.0000}   & \textbf{0.9826} & \textbf{0.9912}\\
\dataset{4}{Arrhythmia ECG}
\multicolumn{1}{p{2.5cm}}{Pan-Tompkins\hfill\cite{pan1985real}} &  0.6676   & 0.4962 & 0.5693\\
\multicolumn{1}{p{2.5cm}}{Bi-RNN\hfill\cite{schuster1997bidirectional}}       &  0.9567   & 0.8156 & 0.8806\\
\multicolumn{1}{p{2.5cm}}{U-Net\hfill\cite{zahid2021robust}}        &   0.9338  &     0.8843 & 0.9084\\
\multicolumn{1}{p{2.5cm}}{MF-PPO}    &  0.9073   & 0.9211 & 0.9141\\
\multicolumn{1}{p{2.5cm}}{MF-SAC}    &  0.8978   & 0.9160 & 0.9068 \\
\multicolumn{1}{p{2.5cm}}{SMF-PPO}    &  0.9446   & 0.9542 & 0.9494\\
\multicolumn{1}{p{2.5cm}}{SMF-SAC}    &  \textbf{0.9543}   & \textbf{0.9567} & \textbf{0.9555}\\
\dataset{4}{Icentia11k}
\multicolumn{1}{p{2.5cm}}{Pan-Tompkins\hfill\cite{pan1985real}} &  0.6321   & 0.5913 & 0.6110\\
\multicolumn{1}{p{2.5cm}}{Bi-RNN\hfill\cite{schuster1997bidirectional}}       &  0.8713   & 0.9204 & 0.8952\\
\multicolumn{1}{p{2.5cm}}{U-Net\hfill\cite{zahid2021robust}}        &   0.9059  &     0.9108 & 0.9083\\
\multicolumn{1}{p{2.5cm}}{MF-PPO}   &  0.9342  & 0.9190 & 0.9265\\
\multicolumn{1}{p{2.5cm}}{MF-SAC}    &  0.9305  & 0.9421 & 0.9363\\
\multicolumn{1}{p{2.5cm}}{SMF-PPO}   &  0.9747  & 0.9802 & 0.9774\\
\multicolumn{1}{p{2.5cm}}{SMF-SAC}   &  \textbf{0.9763}   & \textbf{0.9831} & \textbf{0.9797}\\ \hline
\end{tabular}
\label{tab1}
\end{table}

We first evaluated the best R-peak detection performance. The SMF-PPO and SMF-SAC had episode lengths of 3, i.e., they automatically applied 3 iterative MFs to optimise R-peak detection performance in the last step. The rationale for selecting an episode length of 3 is provided in Section~\ref{sectl}. In contrast, MF-PPO and MF-SAC employed the same RL training framework but with an episode length of 1, corresponding to a single-stage MF. These were included as baselines to represent data-driven optimisation of single-stage MFs. The results are shown in Table~\ref{tab1}, where the performance metrics are given by
\begin{align}
\begin{split}
    &\text{precision} = \frac{\text{TP}}{\text{TP}+\text{FP}}, ~~ \text{recall} = \frac{\text{TP}}{\text{TP}+\text{FN}} \\
    &\text{F-1} = \frac{\text{TP}}{\text{TP} + 0.5 \times(\text{FP} + \text{FN})}
\end{split}
\label{eqmetrics}
\end{align}

Table~\ref{tab1} shows that SMF-SAC consistently achieved the highest precision, recall, and F-1 scores on all datasets.
Across all datasets, the Pan-Tompkins method achieved F-1 scores of $0.62$ or lower, caused by its reliance on historical R-R intervals, which are misleading in arrhythmia ECG signals with irregular R-R intervals. The DL-based methods, Bi-RNN and U-Net, underperformed SMFs. Notably, their performance fell below that of the non-sequential MF-PPO and MF-SAC, highlighting the advantage of the proposed RL paradigm, which directly optimises R-peak detection, over the existing DL paradigm, which relies on minimising proxy loss functions. 
We also observed that MF-PPO and MF-SAC with episode lengths of 1 underperformed SMF-PPO and SMF-SAC with episode lengths of 3, providing empirical evidence that the sequential application of MFs can overcome the inherent limitations in single-stage MF.

\begin{figure}[t]
\centerline{\includegraphics[width=84mm]{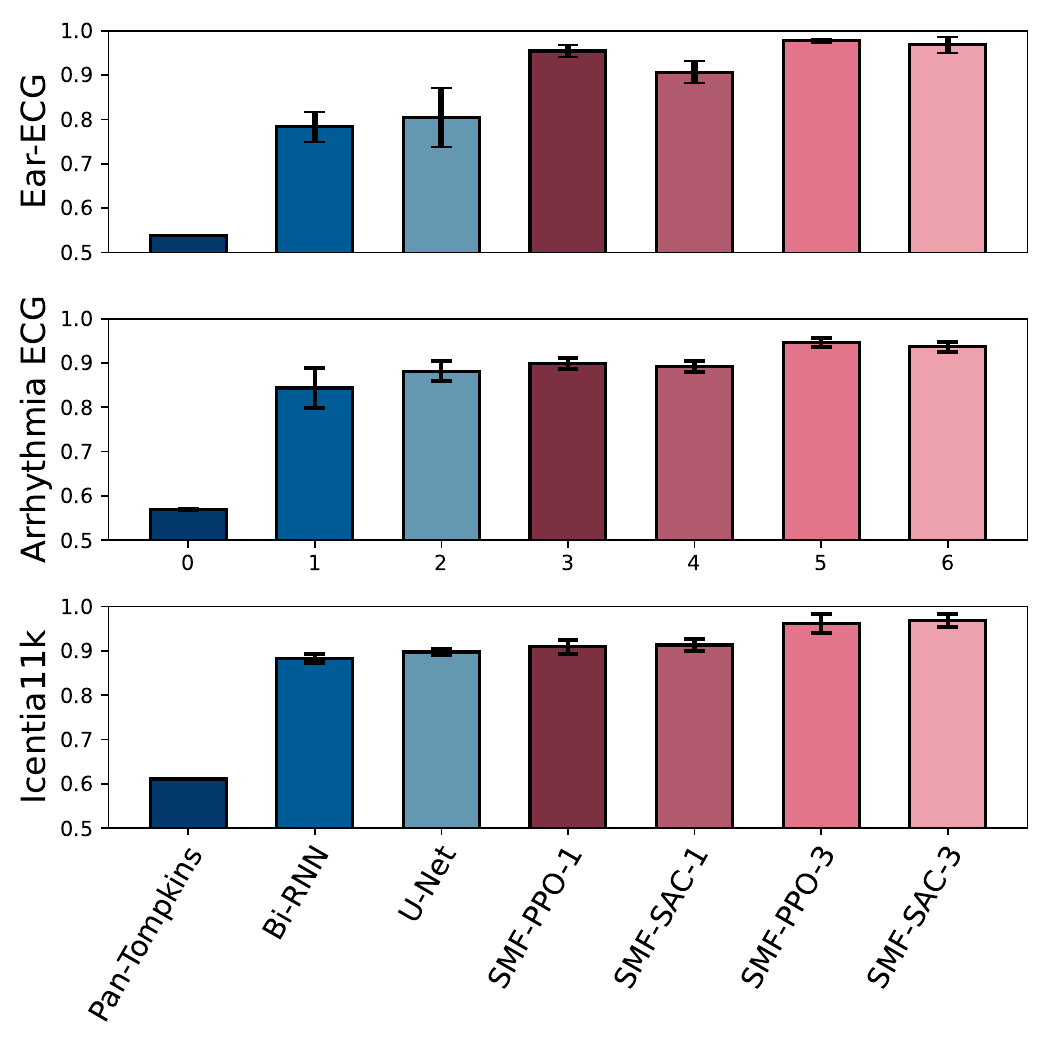}}
\caption{Average F-1 scores of R-peak detection, with the main bars as means and the error bars as standard deviations (where applicable).}
\label{figstd}
\end{figure}

To evaluate the statistical significance of SMF’s performance gains, Fig.~\ref{figstd} presents the average test set F-1 scores of the proposed SMF methods and the baseline methods, where the means and standard deviations were obtained by conducting five independent runs for each method using different random seeds. Observe that SMF consistently achieved the highest average F-1 scores, with low standard deviations compared to the DL-based methods. To show that the advantage of SMF-PPO and SMF-SAC was statistically significant compared to the baselines, we performed t-tests and report the p-values in Table~\ref{tabptest}. We observed that the difference between SMF-SAC and MF-PPO on the ear-ECG dataset is not statistically significant ($p=0.254$), which can be attributed to the inherently higher variance of the SAC algorithm. This indicates that the performance gap between SMF and the non-sequential MF can vary by algorithm. Nevertheless, the sequential SMFs consistently outperformed their non-sequential counterparts, trained using the same RL algorithm. Specifically, SMF-PPO significantly outperformed MF-PPO ($p=0.011$ for ear-ECG and $p=0.0003$ for arrhythmia ECG), and SMF-SAC significantly outperformed MF-SAC ($p=0.004$ for ear-ECG and $p=0.001$).

\begin{table}[t]
\caption{The p-values in t-tests comparing SMF with the baselines.}
\setlength{\tabcolsep}{8pt}
\centering
\begin{tabular}{lcccc}
\hline
\multicolumn{1}{l}{}          & Bi-RNN & U-Net & MF-PPO & MF-SAC \\ \hline
\dataset{5}{Ear-ECG}
\multicolumn{1}{l}{SMF-PPO} & 0.000  & 0.001 & 0.011     & 0.001     \\
\multicolumn{1}{l}{SMF-SAC} & 0.000  & 0.002 & 0.254     & 0.004     \\ 
\dataset{5}{Arrhythmia ECG}
\multicolumn{1}{l}{SMF-PPO} & 0.002  & 0.001 & 0.000     & 0.000     \\
\multicolumn{1}{l}{SMF-SAC} & 0.004  & 0.002 & 0.002     & 0.001     \\ 
\dataset{5}{Icentia11k}
\multicolumn{1}{l}{SMF-PPO} & 0.000  & 0.000 & 0.004     & 0.005    \\
\multicolumn{1}{l}{SMF-SAC} & 0.000  & 0.000 & 0.001     & 0.001     \\ \hline
\end{tabular}
\label{tabptest}
\end{table}

Our previous work has demonstrated that SMF neural networks can run in real time on smartphones~\cite{zylinski2023hearables}. In Table~\ref{tab2}, we compare the average time required by SMF and the baseline methods to process a 1.25 s ECG segment. All methods were averaged over 10 runs across all test segments from the arrhythmia ECG dataset. All experiments were conducted on an Ubuntu 22.04 system with an Intel Core i7-13850HX CPU and an Nvidia RTX 3500 Ada GPU. Being faster than the widely used, real-time Pan-Tompkins method, SMF met the requirement for real-time R-peak detection. Although SMF was slightly slower than U-Net due to its sequential nature, it outperformed the Bi-RNN baseline, as its CNN-based architecture enables greater parallelisation than recurrent models. Despite running two additional MF steps, SMF was only approximately $40\%$ slower than MF. This is because the overall inference cost is dominated by per-segment operations (e.g., data loading and peak finding), rather than the policy forward passes and filtering steps.

\begin{table}[t]
\caption{Average processing time of a 1.25\,s ECG Segment.}
\centering
\setlength{\tabcolsep}{8pt}
\begin{tabular}{p{1.9cm} c | p{1.9cm} c}
\hline
\multicolumn{2}{c|}{Baselines} & \multicolumn{2}{c}{MF methods} \\
\hline
Method & Time (ms) & Method & Time (ms) \\
\hline
Pan-Tompkins & 4.423 & MF-PPO  & 0.970 \\
Bi-RNN       & 2.421 & MF-SAC  & 0.964 \\
U-Net        & 0.987 & SMF-PPO & 1.403 \\
             &       & SMF-SAC & 1.372 \\
\hline
\end{tabular}
\label{tab2}
\end{table}

\begin{figure*}[t]
\centerline{\includegraphics[width=\textwidth]{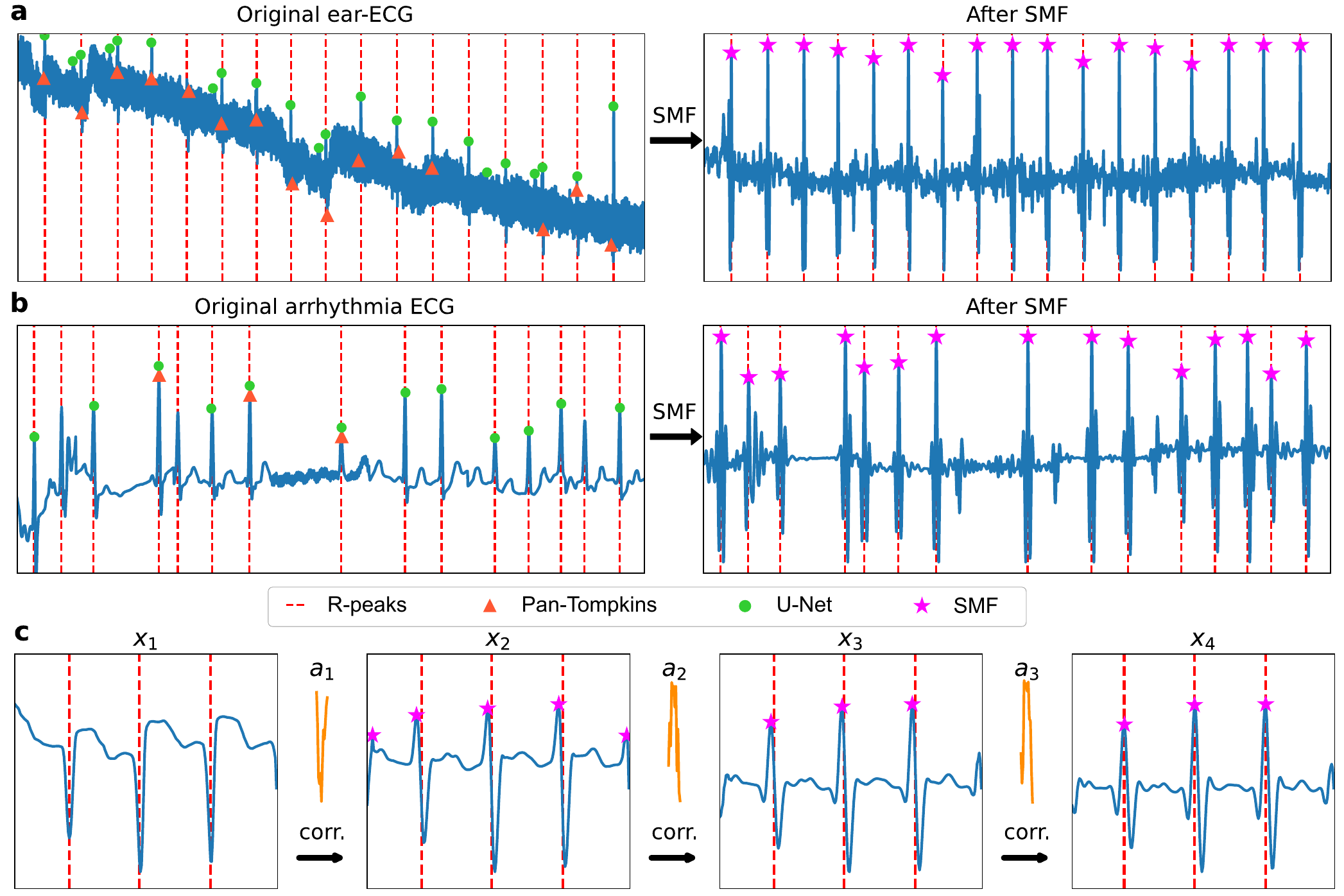}}
\caption{Comparison of the proposed SMF method (SMF-SAC) with the widely used Pan–Tompkins algorithm and the state-of-the-art U-Net on example ECG segments: (\textbf{a}) a noisy ear-ECG section containing numerous false peaks caused by non-cardiac artefacts; (\textbf{b}) an arrhythmia ECG section with varying R–R intervals, which challenge rule-based approaches such as Pan–Tompkins. In (\textbf{c}), we show SMF procedures applied to an arrhythmia ECG segment with inverted R-peaks due to improper recording setup. SMF automatically corrects the inversion by applying signal-aware templates, which remain interpretable (e.g., $a_1$ corresponds to the inverted R-peak morphology). Note that SMF is optimized for R-peak detection; as such, the SMF-transformed signals should be interpreted as task-relevant representations rather than faithful reconstructions of the original waveform.}
\label{figvis}
\end{figure*}

\begin{figure}[t]
\centerline{\includegraphics[width=\textwidth]{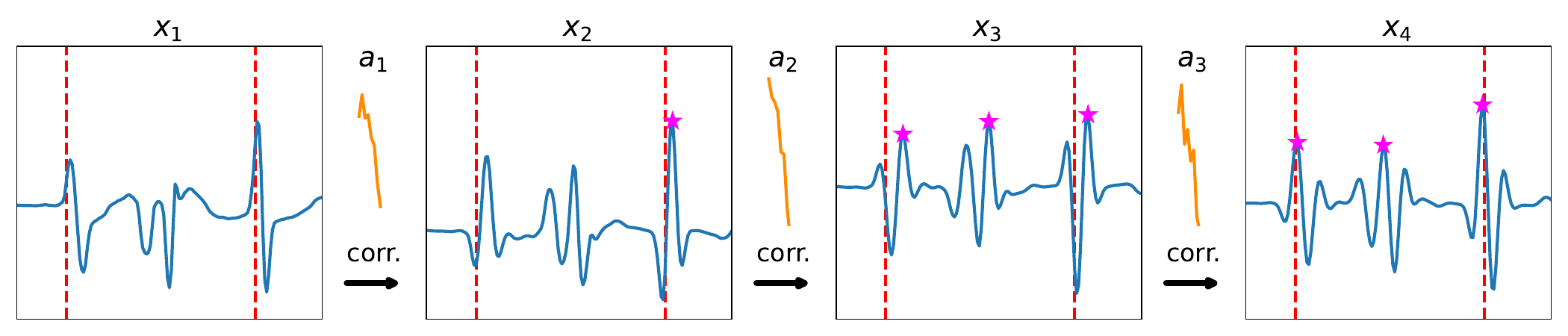}}
\caption{Visualization of an SMF failure case, with the red dashed lines denoting true R-peaks and the pink stars denoting the detected R-peaks. In this case, several transient high-amplitude noise artifacts generate a structured, QRS-like waveform. We note that such structured artifacts are rare in practice, and the same failure was also observed for the baseline methods for this segment.}
\label{figfail}
\end{figure}

\subsection{Visualizing the Robustness}

This work focuses on R-peak detection in the Hearable setting, where a robust ECG R-peak detection method must handle prominent false artefact peaks and ECGs caused by varying patient physiology and recording devices. Fig.~\ref{figvis}a shows the performance of SMF (SMF-SAC) for a noisy ear-ECG section, where false peaks had even greater prominence than the true R-peaks. Nevertheless, SMF localises all true R-peaks. The raw ear-ECG signal also exhibited a baseline drift (see the descending trend in Fig.~\ref{figvis}a), which was effectively removed by SMF. Fig.~\ref{figvis}b shows the performance of SMF on an arrhythmia ECG section, with a large variance in R-R intervals. The rule-based Pan-Tompkins method identified only 3 out of the 15 R-peaks, as its decision rule relies on historical R-R intervals. In contrast, SMF iteratively refined the signal to high quality, eliminating the need for decision rules based on historical R-R intervals. Fig.~\ref{figvis}c shows the complete workflow of SMF on an arrhythmia ECG segment with inverted R-peaks caused by the misplacement of the recording ECG leads. Despite the inversions, SMF successfully localised the R-peaks by generating a series of MF templates with large negative deflections that aligned with the inverted R-peaks. The results also highlight the interpretability of SMF. Compared to the MF templates for healthy patients in Fig.~\ref{figsche}, which exhibited more prominent positive deflections, the templates for arrhythmia patients in Fig.~\ref{figvis}c featured more prominent negative deflections, which could serve as useful indicators for diagnosing cardiovascular anomalies~\cite{libby2005pathophysiology}. 
In Fig.~\ref{figfail}, we show a representative failure case of SMF, in which several transient high-amplitude noise artifacts generated a structured, QRS-like waveform that led to an incorrect detection. We note that such structured artifacts are rare in practice, and the same failure was also observed with the baseline methods for this segment. Practically, the example in Fig.~\ref{figfail} shows how the interpretability of SMF can serve as an indicator of the reliability of R-peak detections. In this example, the wider-than-usual peaks in the learned filters do not align well with plausible QRS-like patterns; accordingly, the corresponding detection results should be treated with caution.

\subsection{Sensitivity Analysis}
We now provide further insights into the proposed SMF method by ablating its design choices and conducting cross-dataset and cross-subject evaluations.

\subsubsection{Impact of Sequential Applications of MFs}
\label{sec:ressmf}

\begin{figure*}[t]
\centerline{\includegraphics[width=\textwidth]{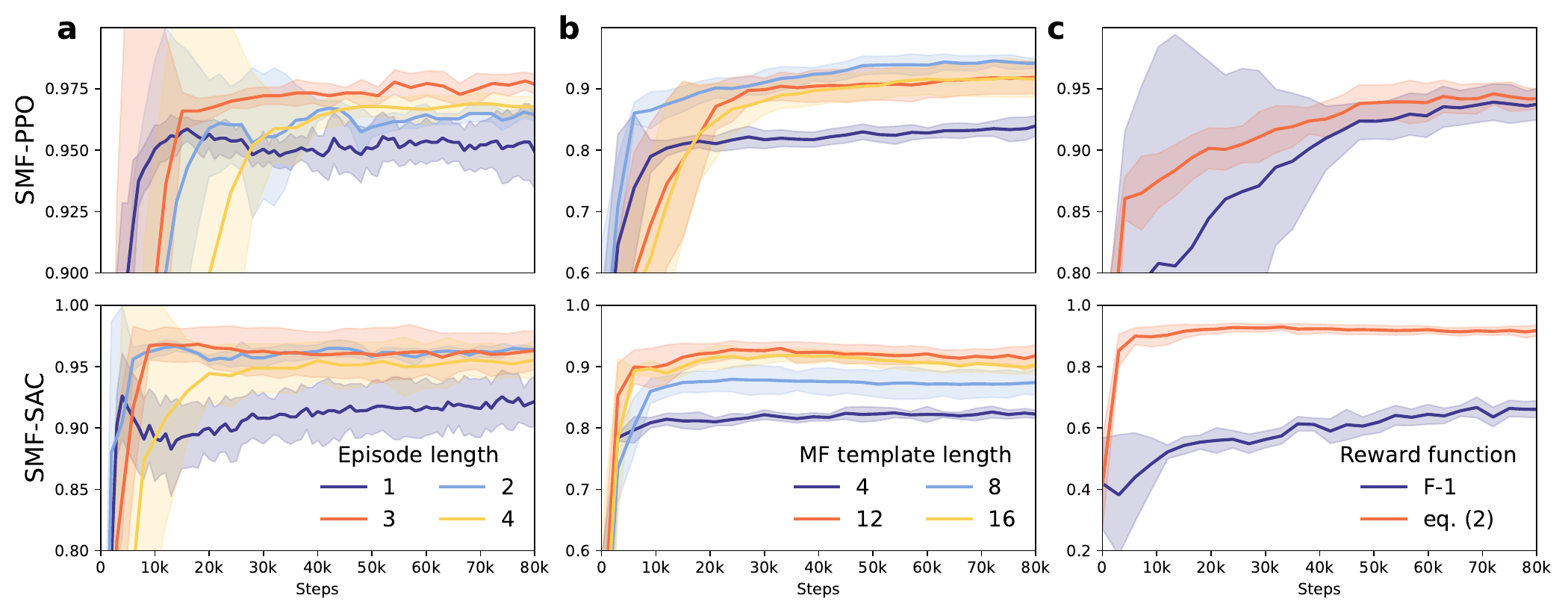}}
\caption{Sensitivity of the SMF method to parameter changes, evaluated by the F-1 scores at different training stages in the ear-ECG environment. The solid lines denote the means, and the shaded areas denote the standard deviations in five independent runs. (\textbf{a}) SMFs with different episode lengths, i.e., the number of MF iterations applied, in the ear-ECG dataset. (\textbf{b}) SMFs with different template lengths in the arrhythmia ECG dataset. (\textbf{c}) SMFs with different reward functions in the arrhythmia ECG dataset.}
\label{figablation}
\end{figure*}

A key insight of this work is that iteratively applying MFs can overcome the inherent limitation of single-stage MFs. To verify this, we trained SMFs with varying episode lengths, i.e., the number of iterative MFs applied, and evaluated their R-peak detection performance (Fig.~\ref{figablation}a). The non-sequential MF-PPO and MF-SAC with episode length 1 achieved average F-1 scores of 0.95 and 0.91, respectively. As the episode length increased, SMFs achieved higher F-1 scores on the test set, with the optimal episode length being 3, where the average F-1 scores were 0.98 and 0.97 for SMF-PPO and SMF-SAC, respectively. This demonstrates that the iterative application of MFs can overcome the limitation of non-sequential MFs. When the episode length was 4, the improvement was marginal for SMF-SAC and even negative for SMF-PPO. 
We also observed that, as the episode length increased, SMFs required more training iterations to converge, with only a modest increase for episode lengths $\leq 3$ and a more pronounced increase at length $=4$.
These were expected, as the sequential decision-making nature of the SMF problem means that the number of samples needed to derive an effective policy grows exponentially with the increase in episode length. Therefore, it became increasingly difficult to learn a stable policy as episode lengths increased.

Overall, on the cost side: (i) within a practical episode length ($<4$ steps), SMF converged in a comparable number of training iterations to its MF counterpart; thus, increasing the episode length in this regime does not substantially increase training difficulty. (ii) the inference cost is dominated by per-segment operations (e.g., data loading and peak-finding), while each SMF step involves only a lightweight policy forward pass and filtering (See Table~\ref{tab2}). As a result, SMFs required only a few milliseconds to process a 1.25~s ECG segment, comfortably meeting real-time constraints for edge deployment. On the benefit side, when comparing like-for-like training algorithms, the sequential variants consistently achieved higher mean performance than their non-sequential counterparts (SMF-PPO vs.\ MF-PPO and SMF-SAC vs.\ MF-SAC, both with $p<0.005$). Taken together, for reasonable episode lengths, the sequential RL formulation provides measurable performance gains at a modest additional training and inference cost, supporting its practical value.

\subsubsection{Effect of Template Length}
\label{sectl}

We explored how MF template length affected SMF performance. We experimented with template lengths of 4, 8, 12, and 16 samples in the arrhythmia ECG dataset (Fig.~\ref{figablation}b). The results show that a template length of 4 yielded the worst performance for both SMF-PPO and SMF-SAC, as such a short template failed to capture distinctive R-peak patterns. Since the QRS pattern around R-peaks lasts about 0.08 seconds (16 samples) in adults, longer MF templates improve correlation with true peaks while reducing artefact correlation. However, increasing template length also exponentially expands the action space, making the sequential decision-making problem more challenging.
This effect was most evident when the template length was 16, where the policy search space becomes substantially larger. In this setting, SMF-PPO exhibited unstable training and a lower converged F-1 score, indicating less effective exploration by the PPO algorithm, whereas SMF-SAC, benefiting from entropy-regularized exploration, achieved robust policy search behaviour and higher converged performance.
The optimal template length was 8 for SMF-PPO and 12 for SMF-SAC. The SMF-SAC performed better with longer templates due to its higher sample efficiency, as it leverages all past transitions, whereas SMF-PPO learns only from transitions generated by the current policy. Additionally, SMF-SAC's entropy regularisation promotes exploration in a broader state-action space, resulting in more stable training and enhanced performance.

\subsubsection{Influence of Reward Function Design} 

We examined the impact of the reward function design on SMF performance (Fig.~\ref{figablation}c). For all other experiments in this section, SMF utilised the reward function in \eqref{eqreward}. To quantify the effect of reward function design, we compared this reward design with another straightforward scheme, where the F-1 score in \eqref{eqmetrics} was used as the reward. For SMF-PPO, although training with the F-1 score as the reward was slower and less stable, the final converged F-1 was comparable to that obtained using the reward from \eqref{eqreward}. For SMF-SAC, using the F-1 score as the reward significantly hindered performance. As the lower panel of Fig.~\ref{figablation}c shows, SMF-SAC with the F-1 score as the reward function achieved a test set F-1 score of only around 0.6, which was lower (by $>0.35$) than that achieved with the reward from \eqref{eqreward}.
While using the F-1 score as the reward reflects the general objective of reducing FP and FN, it does not differentiate the impact of each type of error straightforwardly. In contrast, the reward function $10\mathrm{TP} - 5\mathrm{FP} - 5\mathrm{FN}$ linearly assigns different penalties to FP and FN, enabling the agent to adjust for each type of error and facilitating easier learning in the environment.

\subsubsection{Peak-finding Parameters}
To examine the sensitivity of SMF to the peak-finding parameters used in~\eqref{eqreward}, we conducted an ablation study in which the height and distance parameters of the SciPy find\_peaks() function were varied. Table~\ref{tab5} reports the F-1 scores of SMF-SAC on the arrhythmia ECG dataset under different parameters. We observed that the performance of SMF is stable across reasonable parameter choices, indicating that the performance gain of SMF was not driven by a specific parameter configuration.

\begin{table}[t]
\caption{F-1 under different peak-finding parameters.}
\setlength{\tabcolsep}{8pt}
\centering
\begin{tabular}{@{\hskip4pt}c@{\hskip6pt}c|cccc}
\hline
\multicolumn{2}{c|}{\multirow{2}{*}{}}                                         & \multicolumn{4}{c}{Distance / samples}  \\
\multicolumn{2}{c|}{}                                                          & 20    & 30    & 40    & 50    \\ \hline
\multirow{4}{*}{\rotatebox[origin=c]{90}{Height}} 
& 0.3 & 0.951 & 0.951 & 0.952 & 0.952 \\
& 0.4 & 0.953 & 0.956 & 0.956 & 0.954 \\
& 0.5 & 0.954 & 0.956 & 0.956 & 0.954 \\
& 0.6 & 0.942 & 0.943 & 0.943 & 0.943 \\ \hline
\end{tabular}
\label{tab5}
\end{table}

\begin{table}[t]
\caption{Cross-dataset evaluation on R-peak detection.}
\setlength{\tabcolsep}{4pt}
\centering
\begin{tabular}{lccc}
\hline
\multicolumn{1}{c}{} & Precision & Recall & F-1 score \\\hline
\dataset{4}{Icentia11k $\rightarrow$ arr. (vs. arr. $\rightarrow$ arr.)}

\multicolumn{1}{p{2cm}}{Bi-RNN\hfill\cite{schuster1997bidirectional}}
& 0.944 ($\downarrow 0.013$) & 0.771 ($\downarrow 0.044$) & 0.849 ($\downarrow 0.032$) \\

\multicolumn{1}{p{2cm}}{U-Net\hfill\cite{zahid2021robust}}
& 0.921 ($\downarrow 0.013$) & 0.870 ($\downarrow 0.014$) & 0.895 ($\downarrow 0.013$) \\

\multicolumn{1}{p{2cm}}{MF-PPO}
& 0.901 ($\downarrow 0.007$) & 0.917 ($\downarrow 0.004$) & 0.909 ($\downarrow 0.005$) \\

\multicolumn{1}{p{2cm}}{MF-SAC}
& 0.911 ($\uparrow 0.013$) & 0.893 ($\downarrow 0.023$) & 0.902 ($\downarrow 0.005$) \\

\multicolumn{1}{p{2cm}}{SMF-PPO}
& 0.944 ($\downarrow 0.001$) & 0.932 ($\downarrow 0.022$) & 0.938 ($\downarrow 0.011$) \\

\multicolumn{1}{p{2cm}}{SMF-SAC}
& \textbf{0.954} ($\downarrow 0.000$) & \textbf{0.940} ($\downarrow 0.017$) & \textbf{0.947} ($\downarrow 0.009$) \\
\multicolumn{4}{l}{{\footnotesize\hspace{0.2em}* \textit{arr.} represents the arrhythmia ECG dataset.}}\\
\hline
\end{tabular}
\label{tab3}
\end{table}

\subsubsection{Generalizability}

To assess the generalizability of the proposed SMF framework, we conducted a cross-dataset evaluation, in which the SMF and all baseline methods were trained on the Icentia11k dataset and evaluated on the arrhythmia ECG dataset. The results are summarized in Table~\ref{tab3}, where the values in parentheses indicate the performance change relative to in-dataset evaluation. We observed that the MF methods suffered a much smaller performance drop compared to the DL baselines. This is because MFs were specifically regularized to match the QRS complex, whose morphology is relatively stable across different subjects and recording devices. In contrast, DL-based methods were not guided by such domain knowledge and therefore tend to overfit to device- and subject-specific artifacts in the training data.

\begin{table}[t]
\caption{Per-subject evaluation on the ear-ECG dataset with remaining subjects as the training set.}
\setlength{\tabcolsep}{4pt}
\centering
\begin{tabular}{lccccccc}
\hline
\multicolumn{1}{c}{} & \#1 & \#2 & \#3 & \#4 & \#5 & \#6 & \#7\\\hline
\dataset{8}{U-Net~\textnormal{\cite{zahid2021robust}}}
precision & 0.864 & 0.796 & 0.845 & 0.816 & 0.893 & 0.786 & 0.827 \\
recall    & 1.000 & 0.953 & 0.906 & 0.955 & 0.911 & 0.976 & 0.835 \\
F-1 score & 0.927 & 0.868 & 0.874 & 0.880 & 0.902 & 0.871 & 0.831 \\
\dataset{8}{SMF-PPO}
precision & 1.000 & 0.980 & 0.990 & 0.990 & 0.981 & 0.981 & 0.971 \\
recall    & 1.000 & 0.990 & 0.990 & 0.990 & 1.000 & 0.990 & 0.990 \\
F-1 score & 1.000 & 0.985 & 0.990 & 0.990 & 0.990 & 0.986 & 0.981 \\
\dataset{8}{SMF-SAC}
precision & 1.000 & 0.980 & 0.991 & 0.990 & 0.990 & 0.990 & 0.981 \\
recall    & 1.000 & 0.990 & 1.000 & 0.990 & 1.000 & 0.990 & 0.990 \\
F-1 score & 1.000 & 0.985 & 0.995 & 0.990 & 0.995 & 0.990 & 0.986\\
\hline
\end{tabular}
\label{tab4}
\end{table}

To mimic real-world deployment, where cross-subject robustness is essential, we performed a leave-one-subject-out evaluation. The results are summarized in Table~\ref{tab4}, where each of the seven subjects in the ear-ECG dataset was held out for testing, while the models were trained on the remaining six subjects. We also included the strongest DL baseline, the U-Net, for comparison. We observed that, when deployed to unseen subjects, the DL baseline performance varied significantly (the F-1 score differed by nearly 0.1 between subjects 1 and 7), whereas the proposed SMF methods remained highly consistent (the maximum F-1 difference across subjects is below 0.02). This further demonstrated the robustness of the proposed approach: by explicitly matching the physiologically meaningful QRS morphology, SMF learns subject-invariant cardiac features rather than subject-specific artifacts, such as motion artifacts or baseline wander. Overall, the proposed SMF algorithms maintain strong performance and consistently outperform the baselines despite changes in the recording device and subject at deployment, indicating strong generalizability.

\begin{figure}[t]
\centerline{\includegraphics[width=84mm]{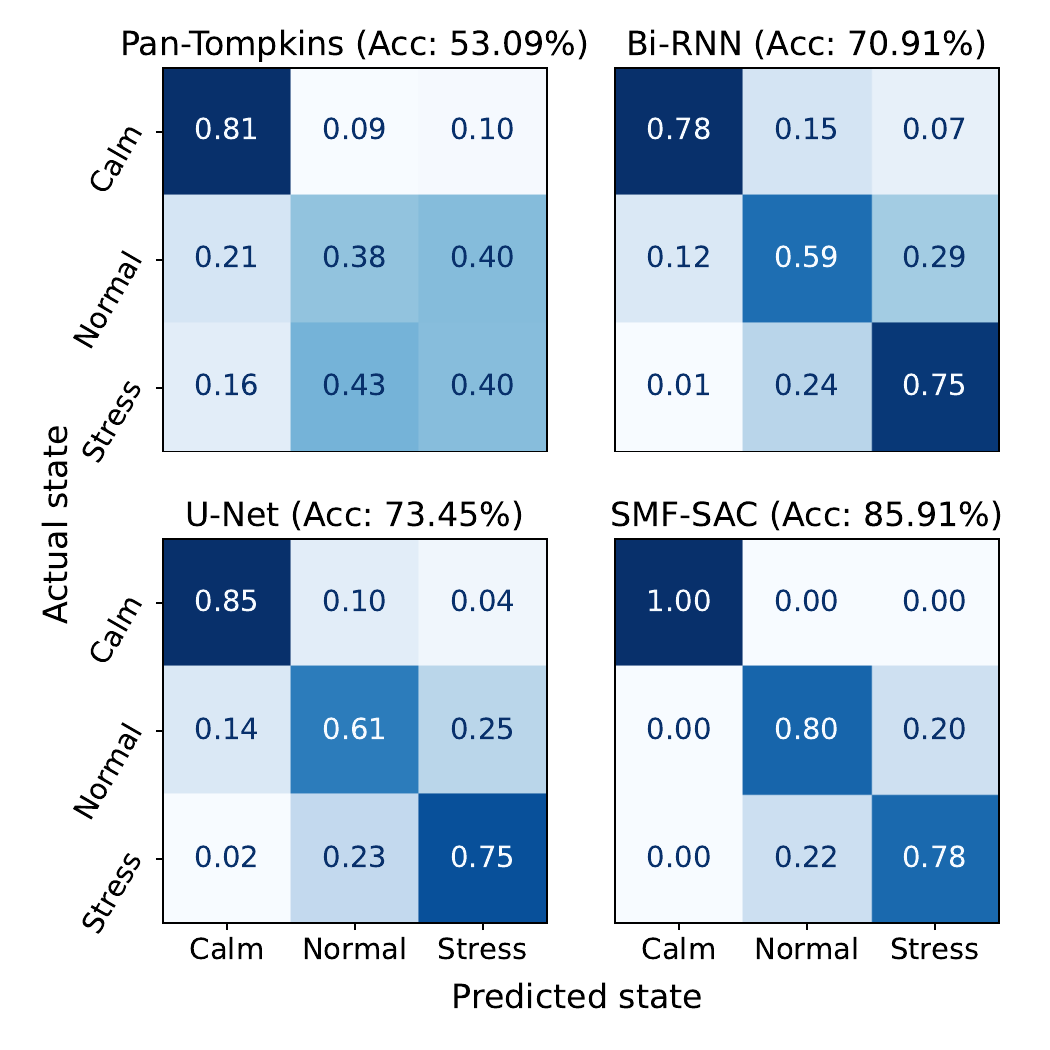}}
\caption{Physiological state classification based on R-peaks extracted in ear-ECG recordings. For each method, the extracted R-peaks were used to compute features for training random forest classifiers. The average accuracy and normalised confusion matrices were obtained over 100 MCCVs using a train-test split of 70:30\%.}
\label{figconfmap}
\end{figure}

\subsection{Physiological State Classification}

As mentioned in Section~\ref{sec:introduction}, rich physiological information can be extracted from R-peaks in ear-ECG. To validate this, we performed a physiological state classification task using the ear-ECG signal. The classification setup was similar to that in~\cite{tian2023hearables}, where three states were considered: calm, during which the subject performed controlled deep breathing; normal, where the subject remained seated and still; and stress, during which the subject solved mental exercises. Each state was recorded for 300 seconds and split into 25-second sections, resulting in a total of 36 sections (12 sections per state).

For both SMF and baseline methods, R-peaks were first localised, followed by the extraction of five widely used Heart Rate Variability (HRV) features: RMSSD, SDNN, HR, LF, HF, and LF/HF (for details on calculating these features, please refer to the summary in~\cite{kim2018stress}). These features were then used to train random forest classifiers. To evaluate classification performance, we computed the average accuracy and normalised confusion matrices over 100 Monte Carlo Cross-Validation (MCCV) runs using a 70:30\% train-test split (Fig.~\ref{figconfmap}). The results showed that features derived from SMF-SAC achieved significantly higher classification accuracy compared to baseline methods. For the calm state, SMF-extracted R-peaks led to perfect classifications, a result that the DL-based methods could not achieve. This highlights the strong potential of SMF for physiological state monitoring on the edge.

\section{Conclusion}
\label{secconc}
Beyond the improved convenience in setup and suitability for prolonged recordings, the rise of edge signal acquisition devices has also created a strong demand for robust, explainable target pattern localisation methods that support trustworthy decision-making. A prominent example is the ear-ECG signals with critically low Signal-to-Noise Ratio, where the reliable localisation of R-peaks can greatly enhance cardiac monitoring and diagnosis. This work addresses this challenge by introducing the Sequential Matched Filter (SMF), which leverages a Reinforcement Learning (RL) agent to design signal-specific filter sequences for robust and interpretable pattern localisation. The RL agent of SMF employs lightweight neural network architectures that are suitable for edge deployment.
When evaluated on three challenging real-world ECG datasets, SMF outperformed state-of-the-art DL baselines and non-sequential MFs optimized using the same RL algorithm.
At the same time, it remains fully interpretable by revealing key signal patterns (e.g., the QRS patterns in ECG) at each step, thereby supporting trustworthy clinical decision-making and enabling the identification of cardiac abnormalities or sensor misplacement. Moreover, we empirically demonstrate that SMF’s improved localisation performance directly enables reliable physiological state classification on the edge. An intriguing conclusion is that SMF provides a robust and interpretable digital filter design framework applicable to edge signal processing tasks beyond the biomedical domain.

\backmatter

\section*{Acknowledgments}
The study protocol for ear-ECG recordings was approved by the Imperial College London Ethics Committee (JRCO 20IC6414) on July 5, 2021.
NM is supported by the UKRI Centre for Doctoral Training in Artificial Intelligence for Healthcare (EP/S023283/1).

\section*{Declarations of Conflict of Interest}

The authors declared that they have no conflicts of interest to this work.

\bibliography{ref}
\end{document}